\documentclass[table, sigconf]{acmart}

\AtBeginDocument{%
  \providecommand\BibTeX{{%
    \normalfont B\kern-0.5em{\scshape i\kern-0.25em b}\kern-0.8em\TeX}}}

\copyrightyear{2024}
\acmYear{2024}
\setcopyright{rightsretained}
\acmConference[CSCW Companion '24]{Companion of the 2024 Computer-Supported Cooperative Work and Social Computing}{November 9--13, 2024}{San Jose, Costa Rica}
\acmBooktitle{Companion of the 2024 Computer-Supported Cooperative Work and Social Computing (CSCW Companion '24), November 9--13, 2024, San Jose, Costa Rica}\acmDOI{10.1145/3678884.3687147}
\acmISBN{979-8-4007-1114-5/24/11}
\acmDOI{10.1145/3678884.3687147}

\hypersetup{breaklinks=true} 
\usepackage{hyperref}
\usepackage{breakurl}
\begin{document}

\title[The Human Factor in AI Red Teaming]{The Human Factor in AI Red Teaming: Perspectives from Social and Collaborative Computing}

\author{Alice Qian Zhang}
\authornote{This work was written while the author was a student at the University of Minnesota.}
\orcid{0009-0005-6407-6981}
\email{aqzhang@andrew.cmu.edu}
\affiliation{%
  \institution{Carnegie Mellon University}
  \city{Pittsburgh}
  \state{Pennsylvania}
  \country{United States}
}

\author{Ryland Shaw}
\orcid{0009-0008-6123-1781}
\email{v-rylandshaw@microsoft.com}
\affiliation{%
  \institution{Microsoft Research}
  \city{New York City}
  \state{New York}
  \country{United States}
}

\author{Jacy Reese Anthis}
\orcid{0000-0002-4684-348X}
\email{anthis@uchicago.edu}
\affiliation{%
  \institution{University of Chicago}
  \city{Chicago}
  \state{Illinois}
  \country{United States}
}

\author{Ashlee Milton}
\orcid{0000-0002-0320-6122}
\email{milto064@umn.edu}
\affiliation{%
  \institution{University of Minnesota}
  \city{Minneapolis}
  \state{Minnesota}
  \country{United States}
}

\author{Emily Tseng}
\orcid{0000-0003-1087-1101}
\email{etseng42@gmail.com}
\affiliation{%
  \institution{Microsoft Research}
  \country{United States}
}

\author{Jina Suh}
\orcid{0000-0002-7646-5563}
\email{jinsuh@microsoft.com}
\affiliation{%
  \institution{Microsoft Research}
  \city{Redmond}
  \state{Washington}
  \country{United States}
}

\author{Lama Ahmad}
\orcid{0009-0009-3017-2499}
\email{lama@openai.com}
\affiliation{%
  \institution{OpenAI}
  \city{San Francisco}
  \state{California}
  \country{United States}
}

\author{Ram Shankar Siva Kumar}
\orcid{0009-0007-8766-3948}
\email{ram.shankar@microsoft.com}
\affiliation{%
  \institution{Microsoft}
  \city{Redmond}
  \state{Washington}
  \country{United States}
}

\author{Julian Posada}
\orcid{0000-0002-3285-6503}
\email{julian.posada@yale.edu}
\affiliation{%
  \institution{Yale University}
  \city{New Haven}
  \state{Connecticut}
  \country{United States}
}

\author{Benjamin Shestakofsky}
\orcid{0000-0003-0797-2729}
\email{bshesta@sas.upenn.edu}
\affiliation{%
  \institution{University of Pennsylvania}
  \city{Philadelphia}
  \state{Pennsylvania}
  \country{United States}
}

\author{Sarah T. Roberts}
\orcid{0000-0001-8953-1470}
\email{sarah.roberts@ucla.edu}
\affiliation{%
  \institution{University of California Los Angeles}
  \city{Los Angeles}
  \state{California}
  \country{United States}
}

\author{Mary L. Gray}
\orcid{0000-0001-9972-6829}
\email{mlg@microsoft.com}
\affiliation{%
  \institution{Microsoft Research}
  \city{Cambridge}
  \state{Massachusetts}
  \country{United States}
}

\renewcommand{\shortauthors}{Zhang et al.}

\begin{abstract}
Rapid progress in general-purpose AI has sparked significant interest in ``red teaming,'' a practice of adversarial testing originating in military and cybersecurity applications. AI red teaming raises many questions about the human factor, such as how red teamers are selected, biases and blindspots in how tests are conducted, and harmful content's psychological effects on red teamers. A growing body of HCI and CSCW literature examines related practices—including data labeling, content moderation, and algorithmic auditing. However, few, if any, have investigated red teaming itself. This workshop seeks to consider the conceptual and empirical challenges associated with this practice, often rendered opaque by non-disclosure agreements. Future studies may explore topics ranging from fairness to mental health and other areas of potential harm. We aim to facilitate a community of researchers and practitioners who can begin to meet these challenges with creativity, innovation, and thoughtful reflection.
\end{abstract}




\maketitle

\section{Introduction}
As machine learning applications---particularly those driven by large language models---have become increasingly widespread, researchers have examined how these technologies may be integrated into our lives while also adhering to responsible artificial intelligence (AI) standards formulated by governments, large technology organizations, and researchers \cite{The_White_House_2023}. Given AI systems' broad application and relatively unpredictable nature, it is challenging for designers and developers to anticipate all possible use cases and consequences. For example, generative AI tools have been shown to reproduce implicit stereotypes about gender and ethnicity \cite{gillespie_generative_2024, bolukbasi_man_2016}. This follows other AI blunders, such as Google's image recognition system tagging photographs of Black people as ``gorillas'' \cite{barr_google_2015} and Microsoft's Tay chatbot engaging in Holocaust denial \cite{schwartz_2016_2019}. These incidents echo a long history of tech companies unintentionally reifying material and representational harms that researchers and policymakers have worked to counteract \cite{barocas_problem_2017}.

The harms caused by these AI systems have inspired numerous efforts at ``red teaming,'' with leading AI companies such as OpenAI, Google, Microsoft, and Anthropic adopting red teams into their responsible AI initiatives. Red teaming has been defined as ``a structured process for probing AI systems and products for the identification of harmful capabilities, outputs, or infrastructural threats'' \cite{FrontierModelForum}. However, what constitutes red teaming changes as the practice evolves, with its definition and methods being shaped by advancements and insights from various fields. For example, Anthropic is using crowd workers to test AI systems by trying to make them generate harmful content \cite{ganguli2022red}. In red teams, users -- from domain experts in cybersecurity and algorithmic fairness to international crowd workers -- take on a pseudo-adversarial role in deliberately producing harmful outputs.

While most red teaming initiatives in AI began recently, there is substantial precedent in other domains. The term was first used to describe military scenario testing, formally established as a practice by the US government during the Cold War \cite{Zenko_Haass_2015}. More recently, it has appeared in computer security \cite{zenko2015red, abbass2011computational, wood2000red}. Today, red teamers use AI systems to generate outputs that are then subject to review–an activity that CSCW scholars have previously examined through the lenses of social computing, content moderation, and data labeling \cite{roberts_behind_2019, gillespie_custodians_2018, gray_ghost_2019}. In this workshop, we hope to draw on scholarship on the history of red teaming to understand and inform the use of red teaming for AI.

Red teaming’s scope differs across applications, and the working practices and occupational hazards related to red teaming are likely to vary widely depending on the labor arrangements in which it is embedded. Red team initiatives rely on contracted experts, permanent employees, volunteers, crowd workers, and end-users \cite{humane_intelligence_SeeAI_DEFCON_AI_Village_2024,ganguli2022red}. Red teamers' identities and organizational contexts may influence AI systems in subtle and unexpected ways. Thus, it is imperative to map the sociotechnical ecology of red teaming--the people doing the work, their methods and means, and their organizational settings. The varying skills users bring to red teaming (e.g., resilience) can also affect their ability to cope with challenges related to productive model assessment and psychological pressures.

Some AI red teamers search for bugs and security risks, but others provoke AI systems to generate content that may be racist, sexist, queerphobic, or have other prejudicial implications. Throughout this process, they expose themselves to the harmful content they help create, with the goal of reducing its availability and effects on end-users of the technology. Repeated exposure to such content has been shown to cause psychological harm to crowd workers and content moderators \cite{steiger2022effects, Michel2018ExContentMS, ruckenstein_re-humanizing_2020, Dwoskin_2019, arsht_2018_human}. We recognize the importance of centering red teamers' well-being in future research \cite{miceli_studying_2022}.

In this workshop, we will explore the evolving landscape of AI red teaming, drawing from both contemporary and historical perspectives. We will explore stakeholder roles in AI red teaming, identifying practitioner needs and addressing worker safety and well-being concerns. Through our discussions and collaborative activities, we aim to achieve two primary outcomes: 1) establish an AI red teaming research network, fostering interdisciplinary collaboration among researchers and practitioners, and 2) collate key insights into an informal post-workshop report for practitioners and researchers.

\section{Workshop Goals and Themes}
This workshop aims to outline the practice of AI red teaming, drawing on historical insights to understand its trajectory and structure. We prioritize understanding the humans involved in AI red teaming and how their roles influence the development of AI systems. Additionally, we seek to leverage past research to address safety concerns and identify academic disciplines and methodologies pertinent to analyzing red teaming practices. We will focus on the following themes:

\begin{enumerate}
    \item \textbf{Conceptualization of Red Teaming} Inspired by Robert Soden and colleagues' ~\cite{soden_time_2021} argument to ground CSCW in history, we aim to understand the trajectory of red teaming as a socio-technical, collaborative practice. This theme invites participants to engage in deeper discussions about red teaming complexities and consider the impact of conducting research in this space ~\cite{olteanu2023responsible}. \textit{What constitutes red teaming, and how has its conceptualization evolved over time? What role does red teaming play within broader frameworks of Responsible AI, and how can decentralized or external approaches contribute to its effectiveness?}
    
    \item \textbf{Labor of Red Teaming} This theme explores the human aspects of AI red teaming, investigating stakeholders involved in the practice and their impact on shaping AI systems to inform future practices and policies. By examining the labor arrangements and power dynamics involved in red teaming practices (e.g., inequities in organizational practices of tech labor ~\cite{shestakofsky2017working}), we seek to uncover historical parallels and contemporary methodologies that illuminate red teamers' roles and operational frameworks.
    \textit{What can historical precedents teach us about red teaming as a labor practice? How can we employ diverse methodologies to investigate red teamers' labor structures, including recruitment procedures and institutional commitments?} 
    
    \item \textbf{Well-being of and Harms Against Red Teamers} Building on the theme of labor, this theme focuses on the safety and well-being of red teamers. We will identify strategies and interventions to mitigate potential harms from exposure to harmful content during red teaming activities. By addressing these critical concerns and integrating recommendations to prioritize worker well-being (e.g., ~\cite{pendse2024towards}), we aim to foster a culture of well-being within the AI red teaming community.
    \textit{How can organizations build safeguards and design interventions to protect red teamers from potential harm? How can these strategies be implemented to ensure the safety and well-being of red teamers in their roles?}
\end{enumerate}

\section{Workshop Activities}
We propose a one-day hybrid workshop that participants can attend in person at CSCW 2024 or virtually to include as many perspectives as possible. We will account for conference and meal breaks in the final schedule. We aim to include the following activities in the workshop:

\begin{itemize}
    \item \textbf{Introduction (15 minutes):} Workshop organizers will welcome participants, introduce themselves, and provide an overview of the workshop, including the details of the objectives and planned activities.
    
    \item \textbf{Red Teaming Exercise (45 minutes):} We will have two red teaming exercises led by workshop organizers. First, participants will be grouped by areas of expertise (e.g., fairness~\cite{anthis2024impossibility}, mental health~\cite{pendse2024towards}) accompanied by a workshop organizer per group. Participants will be given a brief walkthrough of prompt injection via the \href{https://gandalf.lakera.ai/}{Gandalf platform} in which the goal is to force the system to share a secret password. Then, each group will be instructed to formulate and input prompts into a popular LLM (e.g., ChatGPT, Claude) that produces harmful system behavior while being mindful of usage policies. Participants will produce a one-page ``report'' of their findings, emulating an open-ended task assigned to expert red teamers.

    \item \textbf{Panel Discussion (1 hour):} To provide further background for this new area of CSCW inquiry, our organizing team includes senior scholars and practitioners from diverse sectors who will speak on challenges in red teaming. These senior workshop organizers will briefly discuss their perspectives on red teaming guided by prepared and live questions.

    \item \textbf{Artifact Development (2 hours):} In this activity, participants will work in small groups to develop preliminary artifacts for later publication. Groups of participants will be no larger than five people, and each will be assigned a predefined workshop theme or theme that surfaced during the workshop. Participants can choose to develop artifacts in one of two ways:
    \begin{itemize}
        \item \textit{Research Agenda:} The first option is to ideate toward a research agenda linking relevant prior research to workshop themes (e.g., organizational practices of tech labor ~\cite{shestakofsky2017working}, prioritizing worker well-being ~\cite{pendse2024towards}, and critically reflecting on research impacts ~\cite{olteanu2023responsible}).
        
        \item \textit{ToolKit:} The second option is to ideate towards a toolkit that may be leveraged by red teaming practitioners relevant to the assigned theme (e.g., for harms, a toolkit might include activities to consider well-being). 
    \end{itemize}
    
    After 90 minutes, groups working on the same theme will discuss and organize their findings in a shared online space (e.g., Miro board).
    
    \item \textbf{Shareouts (1 hour):} To synthesize key findings and common themes across the artifact development activity, each group will have 10-15 minutes to present their findings, while others can provide feedback.

    \item \textbf{Closing Remarks (15 minutes):} An organizer will convene to recap key insights gleaned from the workshop activities. Those interested in refining the artifacts created will have their information collected through a co-authorship interest form. We will emphasize that all contributions made during the workshop will be duly recognized in any resulting publications through acknowledgment if not co-authorship. Following this, there will be an opportunity for additional discussions and to formally conclude the workshop.
\end{itemize}

\section{Hybrid Workshop Logistics}
In line with our workshop's goal of incorporating diverse perspectives, we are fully dedicated to offering a hybrid experience to enhance accessibility. Before the workshop, participants will be asked to bring a device capable of accessing online platforms listed below as these tools will be essential for facilitating the hybrid experience.

\begin{itemize}
    \item \textbf{Website:} The organizers will create a website for the workshop, including information about the workshop, a call for participation, an expression of interest form, the workshop schedule, and any other relevant information.

    \item \textbf{Discord:} We will create a Discord server for the workshop to facilitate participants' interactions before, during, and after the workshop. We will ask participants to introduce themselves and share why they are attending the workshop on the server. The organizers will monitor the server to foster discussions and keep participants engaged.

    \item \textbf{Zoom Video Conferencing:} All activities will be broadcast through Zoom to allow virtual participants to engage in the activities and discussions. The organizers will be on Zoom and will monitor the chat to help facilitate interactions between in-person and virtual participants. We will utilize breakout rooms to pair in-person and virtual participants to promote discussion between modalities.

    \item \textbf{Online Platforms:} To support the sharing and recording of ideas during the discussion activities, we will use Google Documents and Miro to allow workshop participants to take notes. Both of these software are available online, allowing virtual participants to access the materials. 
\end{itemize}

\section{Call for Participants}
We welcome 20-30 academics and practitioners who are working, researching, or interested in red teaming from fields including but not limited to CSCW, AI, HCI, sociology, communications, philosophy, psychology, and labor studies. To express interest in attending, individuals must submit a statement of interest that will include a summary of their motivation for attending the workshop, themes they are interested in exploring, and a short biography via a Google form. The submissions will be reviewed by the workshop organizers and accepted based on the diversity of perspectives, given the focus on bringing together academic and practitioner viewpoints and approaches from diverse domains. We will advertise our workshop through social media and academic mailing lists and reach out to organizations or special interest groups that may be interested in the topic of our workshop.

\section{Workshop Outcomes}
In addition to providing a venue for those already engaged with red teaming, we envision this workshop as an opportunity to highlight red teaming and its relevance to others in CSCW. Thus, the expected outcomes of this workshop include the following:
\begin{itemize}
    \item \textbf{AI Red Teaming Research Network:} We will bring together interdisciplinary researchers and practitioners to critically examine the current state of red teaming and how practitioners can support the humans who perform it. Through the exchange of experiences and ideas, we expect that collaborations will be formed, creating a network of researchers who will continue to work together on the topic.  
    \item \textbf{Synthesis of Workshop Findings:} Building on the collaborative efforts during the workshop activities, we aim to synthesize and publish key findings from the discussions and artifacts developed. We will gather insights and perspectives on AI red teaming practices by discussing developed artifacts. The resulting synthesis will offer valuable insights for practitioners and highlight avenues for further exploration by researchers.
\end{itemize}

\section{Organizing Team}
Our workshop proposal includes a team of researchers experienced in critically examining labor, moderation, mental health and well-being, and RAI red-teaming.

\noindent\textbf{Alice Qian Zhang} is a PhD student in the Human-Computer Interaction Institute at Carnegie Mellon University. Her research explores avenues to support individuals engaging with social computing technologies and AI systems, with a particular emphasis on underrepresented populations and the implications for mental health and well-being.

\noindent\textbf{Ryland Shaw} is a pre-doctoral research assistant at Microsoft Research's Social Media Collective, where he works on questions of AI ethics, technological norms, and sociotechnical systems. He has an MA in Communication from Simon Fraser University and comes from a background in documentary filmmaking. 

\noindent\textbf{Jacy Reese Anthis} is the director of the Sentience Institute, a visiting scholar at the Stanford Institute for Human-Centered Artificial Intelligence (HAI), and a PhD candidate in the sociology and statistics departments at the University of Chicago. Jacy researches machine learning and human-AI interaction, particularly the rise of digital minds and how humanity can work together with highly capable AI systems.

\noindent\textbf{Ashlee Milton} is a PhD candidate in computer science at the University of Minnesota, focusing on human-computer interaction. Their research investigates how information retrieval systems are used by and affect users from marginalized populations from a user perspective to better design these systems to support users' needs and mental well-being. 

\noindent\textbf{Emily Tseng} is a postdoctoral researcher at Microsoft Research. Her work explores how computing technologies mediate individual, interpersonal, and structural harms, and how to create more equitable tech. Emily publishes at top-tier venues in HCI and design (CHI, CSCW), computer security and privacy (USENIX Security), and medicine (JAMA). She earned a Ph.D. in Information Science at Cornell University and a B.A. at Princeton University.

\noindent\textbf{Jina Suh} is a Principal Researcher in the Human Understanding and Empathy group at Microsoft Research. Her work lies at the intersection of technology and human well-being, where she examines the role of technologies, design choices, development practices, and values embedded in them in shifting power dynamics and affecting individual and organizational mental health and well-being. She received her Ph.D. in Computer Science at the University of Washington. 

\noindent\textbf{Lama Ahmad} is a Policy Researcher at OpenAI, leading red teaming and researcher access efforts. Her work focuses on evaluating the socio-technical impact of AI systems on society. Prior to OpenAI, Lama was at Facebook, assessing the impact of social media on elections and democracy. 

\noindent\textbf{Ram Shankar Siva Kumar} founded and leads the AI Red Team at Microsoft and co-authored \textit{Not with a Bug, But with a Sticker: Attacks on Machine Learning Systems and What To Do About Them}~\cite{kumar2023not}. He is also a Tech Policy Fellow at UC Berkeley, wherein his work on adversarial machine learning appeared notably in the National Security Commission on Artificial Intelligence (NSCAI) Final report presented to the United States Congress and the President.

\noindent\textbf{Julian Posada} is an Assistant Professor of American Studies at Yale University and a member of the Yale Law School’s Information Society Project and the Yale Institute for Foundations of Data Science. Their research integrates theories and methods from information science, sociology, and human-computer interaction to examine how technology is developed and used within various historical, cultural, and social contexts.

\noindent\textbf{Benjamin Shestakofsky} is an Assistant Professor of Sociology at the University of Pennsylvania. His research centers on the relationship between work, technology, organizations, and political economy. He is the author of \textit{Behind the Startup: How Venture Capital Shapes Work, Innovation, and Inequality}~\cite{shestakofsky2017working}.

\noindent\textbf{Sarah T. Roberts} is an associate professor at UCLA specializing in Internet and social media policy, infrastructure, politics and culture, and the intersection of media, technology, and society. She is the faculty director of the UCLA Center for Critical Internet Inquiry (C2i2). Informed by feminist Science and Technology Studies perspectives, Roberts is keenly interested in the way power, geopolitics, and economics play out on/via the internet, reproducing, reifying, and exacerbating global inequities and social injustice. 

\noindent\textbf{Mary L. Gray} is a Senior Principal Researcher at Microsoft Research, a Faculty Associate at Harvard University’s Berkman Klein Center for Internet and Society, and a MacArthur Fellow. An anthropologist and media scholar by training, she focuses on how people's everyday uses of technologies transform labor, identity, and human rights. She maintains a faculty position in the Luddy School of Informatics, Computing, and Engineering with affiliations in Anthropology and Gender Studies at Indiana University.

\bibliographystyle{ACM-Reference-Format}
\bibliography{_references}


\begin{thebibliography}{28}


\ifx \showCODEN    \undefined \def \showCODEN     #1{\unskip}     \fi
\ifx \showDOI      \undefined \def \showDOI       #1{#1}\fi
\ifx \showISBNx    \undefined \def \showISBNx     #1{\unskip}     \fi
\ifx \showISBNxiii \undefined \def \showISBNxiii  #1{\unskip}     \fi
\ifx \showISSN     \undefined \def \showISSN      #1{\unskip}     \fi
\ifx \showLCCN     \undefined \def \showLCCN      #1{\unskip}     \fi
\ifx \shownote     \undefined \def \shownote      #1{#1}          \fi
\ifx \showarticletitle \undefined \def \showarticletitle #1{#1}   \fi
\ifx \showURL      \undefined \def \showURL       {\relax}        \fi
\providecommand\bibfield[2]{#2}
\providecommand\bibinfo[2]{#2}
\providecommand\natexlab[1]{#1}
\providecommand\showeprint[2][]{arXiv:#2}

\bibitem[Abbass et~al\mbox{.}(2011)]%
        {abbass2011computational}
\bibfield{author}{\bibinfo{person}{Hussein Abbass}, \bibinfo{person}{Axel Bender}, \bibinfo{person}{Svetoslav Gaidow}, {and} \bibinfo{person}{Paul Whitbread}.} \bibinfo{year}{2011}\natexlab{}.
\newblock \showarticletitle{Computational red teaming: Past, present and future}.
\newblock \bibinfo{journal}{\emph{IEEE Computational Intelligence Magazine}} \bibinfo{volume}{6}, \bibinfo{number}{1} (\bibinfo{year}{2011}), \bibinfo{pages}{30--42}.
\newblock


\bibitem[Anthis et~al\mbox{.}(2024)]%
        {anthis2024impossibility}
\bibfield{author}{\bibinfo{person}{Jacy~Reese Anthis}, \bibinfo{person}{Kristian Lum}, \bibinfo{person}{Michael Ekstrand}, \bibinfo{person}{Avi Feller}, \bibinfo{person}{Alexander D'Amour}, {and} \bibinfo{person}{Chenhao Tan}.} \bibinfo{year}{2024}\natexlab{}.
\newblock \showarticletitle{The Impossibility of Fair LLMs}.
\newblock \bibinfo{journal}{\emph{arXiv preprint arXiv:2406.03198}} (\bibinfo{year}{2024}).
\newblock


\bibitem[Arsht and Etcovitch(2018)]%
        {arsht_2018_human}
\bibfield{author}{\bibinfo{person}{Andrew Arsht} {and} \bibinfo{person}{Daniel Etcovitch}.} \bibinfo{year}{2018}\natexlab{}.
\newblock \showarticletitle{The human cost of online content moderation}.
\newblock \bibinfo{journal}{\emph{Harvard Journal of Law and Technology}}  \bibinfo{volume}{2} (\bibinfo{year}{2018}).
\newblock


\bibitem[Barocas et~al\mbox{.}(2017)]%
        {barocas_problem_2017}
\bibfield{author}{\bibinfo{person}{Solon Barocas}, \bibinfo{person}{Kate Crawford}, \bibinfo{person}{Aaron Shapiro}, {and} \bibinfo{person}{Hanna Wallach}.} \bibinfo{year}{2017}\natexlab{}.
\newblock \showarticletitle{The {Problem} {With} {Bias}: {Allocative} {Versus} {Representational} {Harms} in {Machine} {Learning}}. \bibinfo{address}{Philadelphia, PA}.
\newblock


\bibitem[Barr(2015)]%
        {barr_google_2015}
\bibfield{author}{\bibinfo{person}{Alistair Barr}.} \bibinfo{year}{2015}\natexlab{}.
\newblock \showarticletitle{Google {Mistakenly} {Tags} {Black} {People} as ‘{Gorillas},’ {Showing} {Limits} of {Algorithms}}.
\newblock \bibinfo{journal}{\emph{Wall Street Journal}} (\bibinfo{date}{July} \bibinfo{year}{2015}).
\newblock
\showISSN{0099-9660}
\urldef\tempurl%
\url{http://blogs.wsj.com/digits/2015/07/01/google-mistakenly-tags-black-people-as-gorillas-showing-limits-of-algorithms/}
\showURL{%
\tempurl}


\bibitem[Bolukbasi et~al\mbox{.}(2016)]%
        {bolukbasi_man_2016}
\bibfield{author}{\bibinfo{person}{Tolga Bolukbasi}, \bibinfo{person}{Kai-Wei Chang}, \bibinfo{person}{James~Y Zou}, \bibinfo{person}{Venkatesh Saligrama}, {and} \bibinfo{person}{Adam~T Kalai}.} \bibinfo{year}{2016}\natexlab{}.
\newblock \showarticletitle{Man is to {Computer} {Programmer} as {Woman} is to {Homemaker}? {Debiasing} {Word} {Embeddings}}. In \bibinfo{booktitle}{\emph{30th {Conference} on {Neural} {Information} {Processing} {Systems}}}. \bibinfo{address}{Barcelona}.
\newblock


\bibitem[Dwoskin(2019)]%
        {Dwoskin_2019}
\bibfield{author}{\bibinfo{person}{Elizabeth Dwoskin}.} \bibinfo{year}{2019}\natexlab{}.
\newblock \bibinfo{title}{Inside facebook, the second-class workers who do the hardest job are waging a quiet battle}.
\newblock
\newblock
\urldef\tempurl%
\url{https://www.washingtonpost.com/technology/2019/05/08/inside-facebook-second-class-workers-who-do-hardest-job-are-waging-quiet-battle/}
\showURL{%
\tempurl}


\bibitem[Forum(2024)]%
        {FrontierModelForum}
\bibfield{author}{\bibinfo{person}{Frontier~Model Forum}.} \bibinfo{year}{2024}\natexlab{}.
\newblock \bibinfo{title}{Issue brief: What is red teaming?}
\newblock
\newblock
\urldef\tempurl%
\url{https://www.frontiermodelforum.org/updates/red-teaming/}
\showURL{%
\tempurl}


\bibitem[Ganguli et~al\mbox{.}(2022)]%
        {ganguli2022red}
\bibfield{author}{\bibinfo{person}{Deep Ganguli}, \bibinfo{person}{Liane Lovitt}, \bibinfo{person}{Jackson Kernion}, \bibinfo{person}{Amanda Askell}, \bibinfo{person}{Yuntao Bai}, \bibinfo{person}{Saurav Kadavath}, \bibinfo{person}{Ben Mann}, \bibinfo{person}{Ethan Perez}, \bibinfo{person}{Nicholas Schiefer}, \bibinfo{person}{Kamal Ndousse}, {et~al\mbox{.}}} \bibinfo{year}{2022}\natexlab{}.
\newblock \showarticletitle{Red teaming language models to reduce harms: Methods, scaling behaviors, and lessons learned}.
\newblock \bibinfo{journal}{\emph{arXiv preprint arXiv:2209.07858}} (\bibinfo{year}{2022}).
\newblock


\bibitem[Gillespie(2018)]%
        {gillespie_custodians_2018}
\bibfield{author}{\bibinfo{person}{Tarleton Gillespie}.} \bibinfo{year}{2018}\natexlab{}.
\newblock \bibinfo{booktitle}{\emph{Custodians of the {Internet}: {Platforms}, {Content} {Moderation}, and the {Hidden} {Decisions} {That} {Shape} {Social} {Media}}}.
\newblock \bibinfo{publisher}{Yale University Press}, \bibinfo{address}{New Haven Connecticut}.
\newblock
\showISBNx{978-0-300-17313-0}


\bibitem[Gillespie(2024)]%
        {gillespie_generative_2024}
\bibfield{author}{\bibinfo{person}{Tarleton Gillespie}.} \bibinfo{year}{2024}\natexlab{}.
\newblock \showarticletitle{Generative {AI} and the {Politics} of {Visibility}}.
\newblock \bibinfo{journal}{\emph{Big Data \& Society}} (\bibinfo{year}{2024}).
\newblock


\bibitem[Gray and Suri(2019)]%
        {gray_ghost_2019}
\bibfield{author}{\bibinfo{person}{Mary~L. Gray} {and} \bibinfo{person}{Siddharth Suri}.} \bibinfo{year}{2019}\natexlab{}.
\newblock \bibinfo{booktitle}{\emph{Ghost {Work}: {How} to {Stop} {Silicon} {Valley} from {Building} a {New} {Global} {Underclass}} (\bibinfo{edition}{illustrated edition} ed.)}.
\newblock \bibinfo{publisher}{Harper Business}, \bibinfo{address}{Boston}.
\newblock
\showISBNx{978-1-328-56624-9}


\bibitem[House(2023)]%
        {The_White_House_2023}
\bibfield{author}{\bibinfo{person}{The~White House}.} \bibinfo{year}{2023}\natexlab{}.
\newblock \bibinfo{title}{FACT SHEET: Biden-Harris Administration Secures Voluntary Commitments from Leading Artificial Intelligence Companies to Manage the Risks Posed by AI}.
\newblock
\newblock
\urldef\tempurl%
\url{https://www.whitehouse.gov/briefing-room/statements-releases/2023/07/21/fact-sheet-biden-harris-administration-secures-voluntary-commitments-from-leading-artificial-intelligence-companies-to-manage-the-risks-posed-by-ai/}
\showURL{%
\tempurl}


\bibitem[humane intelligence et~al\mbox{.}(2024)]%
        {humane_intelligence_SeeAI_DEFCON_AI_Village_2024}
\bibfield{author}{\bibinfo{person}{humane intelligence}, \bibinfo{person}{SeeAI}, {and} \bibinfo{person}{DEFCON~AI Village}.} \bibinfo{year}{2024}\natexlab{}.
\newblock
\newblock


\bibitem[Kumar and Anderson(2023)]%
        {kumar2023not}
\bibfield{author}{\bibinfo{person}{Ram Shankar~Siva Kumar} {and} \bibinfo{person}{Hyrum Anderson}.} \bibinfo{year}{2023}\natexlab{}.
\newblock \bibinfo{booktitle}{\emph{Not with a Bug, But with a Sticker: Attacks on Machine Learning Systems and what to Do about Them}}.
\newblock \bibinfo{publisher}{John Wiley \& Sons, Incorporated}.
\newblock


\bibitem[Miceli et~al\mbox{.}(2022)]%
        {miceli_studying_2022}
\bibfield{author}{\bibinfo{person}{Milagros Miceli}, \bibinfo{person}{Julian Posada}, {and} \bibinfo{person}{Tianling Yang}.} \bibinfo{year}{2022}\natexlab{}.
\newblock \showarticletitle{Studying {Up} {Machine} {Learning} {Data}: {Why} {Talk} {About} {Bias} {When} {We} {Mean} {Power}?}
\newblock \bibinfo{journal}{\emph{Proceedings of the ACM on Human-Computer Interaction}} \bibinfo{volume}{6}, \bibinfo{number}{GROUP} (\bibinfo{date}{Jan.} \bibinfo{year}{2022}), \bibinfo{pages}{34:1--34:14}.
\newblock
\urldef\tempurl%
\url{https://doi.org/10.1145/3492853}
\showDOI{\tempurl}


\bibitem[Michel(2018)]%
        {Michel2018ExContentMS}
\bibfield{author}{\bibinfo{person}{Michel}.} \bibinfo{year}{2018}\natexlab{}.
\newblock \showarticletitle{Ex-Content Moderator Sues Facebook, Saying Violent Images Caused Her PTSD - e-traces}.
\newblock
\urldef\tempurl%
\url{https://api.semanticscholar.org/CorpusID:150309904}
\showURL{%
\tempurl}


\bibitem[Olteanu et~al\mbox{.}(2023)]%
        {olteanu2023responsible}
\bibfield{author}{\bibinfo{person}{Alexandra Olteanu}, \bibinfo{person}{Michael Ekstrand}, \bibinfo{person}{Carlos Castillo}, {and} \bibinfo{person}{Jina Suh}.} \bibinfo{year}{2023}\natexlab{}.
\newblock \showarticletitle{Responsible AI Research Needs Impact Statements Too}.
\newblock \bibinfo{journal}{\emph{arXiv preprint arXiv:2311.11776}} (\bibinfo{year}{2023}).
\newblock


\bibitem[Pendse et~al\mbox{.}(2024)]%
        {pendse2024towards}
\bibfield{author}{\bibinfo{person}{Sachin~R Pendse}, \bibinfo{person}{Talie Massachi}, \bibinfo{person}{Jalehsadat Mahdavimoghaddam}, \bibinfo{person}{Jenna Butler}, \bibinfo{person}{Jina Suh}, {and} \bibinfo{person}{Mary Czerwinski}.} \bibinfo{year}{2024}\natexlab{}.
\newblock \showarticletitle{Towards Inclusive Futures for Worker Wellbeing}.
\newblock \bibinfo{journal}{\emph{Proceedings of the ACM on Human-Computer Interaction}} \bibinfo{volume}{8}, \bibinfo{number}{CSCW1} (\bibinfo{year}{2024}), \bibinfo{pages}{1--32}.
\newblock


\bibitem[Roberts(2019)]%
        {roberts_behind_2019}
\bibfield{author}{\bibinfo{person}{Sarah~T. Roberts}.} \bibinfo{year}{2019}\natexlab{}.
\newblock \bibinfo{booktitle}{\emph{Behind the {Screen}: {Content} {Moderation} in the {Shadows} of {Social} {Media}}}.
\newblock \bibinfo{publisher}{Yale University Press}.
\newblock
\showISBNx{978-0-300-24531-8}
\newblock
\shownote{Google-Books-ID: uiCbDwAAQBAJ}.


\bibitem[Ruckenstein and Turunen(2020)]%
        {ruckenstein_re-humanizing_2020}
\bibfield{author}{\bibinfo{person}{Minna Ruckenstein} {and} \bibinfo{person}{Linda Lisa~Maria Turunen}.} \bibinfo{year}{2020}\natexlab{}.
\newblock \showarticletitle{Re-humanizing the platform: Content moderators and the logic of care}.
\newblock  \bibinfo{volume}{22}, \bibinfo{number}{6} (\bibinfo{year}{2020}), \bibinfo{pages}{1026--1042}.
\newblock
\newblock
\shownote{Publisher: Sage Publications Sage {UK}: London, England}.


\bibitem[Schwartz(2019)]%
        {schwartz_2016_2019}
\bibfield{author}{\bibinfo{person}{Oscar Schwartz}.} \bibinfo{year}{2019}\natexlab{}.
\newblock \bibinfo{title}{In 2016, {Microsoft}’s {Racist} {Chatbot} {Revealed} the {Dangers} of {Online} {Conversation} - {IEEE} {Spectrum}}.
\newblock
\newblock
\urldef\tempurl%
\url{https://spectrum.ieee.org/in-2016-microsofts-racist-chatbot-revealed-the-dangers-of-online-conversation}
\showURL{%
\tempurl}


\bibitem[Shestakofsky(2017)]%
        {shestakofsky2017working}
\bibfield{author}{\bibinfo{person}{Benjamin Shestakofsky}.} \bibinfo{year}{2017}\natexlab{}.
\newblock \showarticletitle{Working algorithms: Software automation and the future of work}.
\newblock \bibinfo{journal}{\emph{Work and Occupations}} \bibinfo{volume}{44}, \bibinfo{number}{4} (\bibinfo{year}{2017}), \bibinfo{pages}{376--423}.
\newblock


\bibitem[Soden et~al\mbox{.}(2021)]%
        {soden_time_2021}
\bibfield{author}{\bibinfo{person}{Robert Soden}, \bibinfo{person}{David Ribes}, \bibinfo{person}{Seyram Avle}, {and} \bibinfo{person}{Will Sutherland}.} \bibinfo{year}{2021}\natexlab{}.
\newblock \showarticletitle{Time for {Historicism} in {CSCW}: {An} {Invitation}}.
\newblock \bibinfo{journal}{\emph{Proceedings of the ACM on Human-Computer Interaction}} \bibinfo{volume}{5}, \bibinfo{number}{CSCW2} (\bibinfo{date}{Oct.} \bibinfo{year}{2021}), \bibinfo{pages}{1--18}.
\newblock
\showISSN{2573-0142}
\urldef\tempurl%
\url{https://doi.org/10.1145/3479603}
\showDOI{\tempurl}


\bibitem[Steiger et~al\mbox{.}(2022)]%
        {steiger2022effects}
\bibfield{author}{\bibinfo{person}{Miriah Steiger}, \bibinfo{person}{Timir~J Bharucha}, \bibinfo{person}{Wilfredo Torralba}, \bibinfo{person}{Marlyn Savio}, \bibinfo{person}{Priyanka Manchanda}, {and} \bibinfo{person}{Rachel Lutz-Guevara}.} \bibinfo{year}{2022}\natexlab{}.
\newblock \showarticletitle{Effects of a Novel Resiliency Training Program for Social Media Content Moderators}. In \bibinfo{booktitle}{\emph{Proceedings of Seventh International Congress on Information and Communication Technology: ICICT 2022, London, Volume 4}}. Springer, \bibinfo{pages}{283--298}.
\newblock


\bibitem[Wood and Duggan(2000)]%
        {wood2000red}
\bibfield{author}{\bibinfo{person}{Bradley~J Wood} {and} \bibinfo{person}{Ruth~A Duggan}.} \bibinfo{year}{2000}\natexlab{}.
\newblock \showarticletitle{Red teaming of advanced information assurance concepts}. In \bibinfo{booktitle}{\emph{Proceedings DARPA Information Survivability Conference and Exposition. DISCEX'00}}, Vol.~\bibinfo{volume}{2}. IEEE, \bibinfo{pages}{112--118}.
\newblock


\bibitem[Zenko(2015)]%
        {zenko2015red}
\bibfield{author}{\bibinfo{person}{Micah Zenko}.} \bibinfo{year}{2015}\natexlab{}.
\newblock \bibinfo{booktitle}{\emph{Red Team: How to succeed by thinking like the enemy}}.
\newblock \bibinfo{publisher}{Basic Books}.
\newblock


\bibitem[Zenko and Haass(2015)]%
        {Zenko_Haass_2015}
\bibfield{author}{\bibinfo{person}{Micah Zenko} {and} \bibinfo{person}{Richard Haass}.} \bibinfo{year}{2015}\natexlab{}.
\newblock \bibinfo{title}{“red team: How to succeed by thinking like the enemy”}.
\newblock
\newblock
\urldef\tempurl%
\url{https://www.cfr.org/event/red-team-how-succeed-thinking-enemy}
\showURL{%
\tempurl}


\end{thebibliography}

\end{document}